\documentclass[twocolumn,superscriptaddress, showpacs] {revtex4}
\usepackage{graphicx}
\usepackage{amsmath}
\usepackage{dsfont}
\usepackage{amssymb}

\begin{document}

\title[Short Title]{Dissipative creation of three-dimensional
entangled state in optical cavity via spontaneous emission}
\author{Xiao-Qiang Shao\footnote{E-mail: xqshao@yahoo.com}}
\affiliation{School of Physics, Northeast Normal University
Changchun 130024, People's Republic of China}
\affiliation{Centre for Quantum Technologies, National University of Singapore, 3 Science Drive 2, Singapore 117543}
\author{Tai-Yu Zheng}
\affiliation{School of Physics, Northeast Normal University
Changchun 130024, People's Republic of China}
\author{C. H. Oh}
\affiliation{Centre for Quantum Technologies, National University of Singapore, 3 Science Drive 2, Singapore 117543}
\author{Shou Zhang}
 \affiliation{Department of Physics, College of Science,
Yanbian University, Yanji, Jilin 133002, People's Republic of China}
\begin{abstract}
We present a dissipative protocol to engineer two $^{87}Rb$ atoms into a form of three-dimensional
entangled state via spontaneous emission. The combination of coupling between ground states via microwave fields and dissipation induced by spontaneous emission make the current scheme  deterministic and a stationary entangled state
can always be achieved without state initialization.  Moreover, this scheme can be straightforwardly  generalized to preparation of an $N$-dimensional entangled state in principle.
\end{abstract}
 \maketitle \maketitle
\section{introduction}
For an open quantum system, the dissipation process must be accompanied by entanglement generation, i.e. the populations of quantum states are altered due to entanglement with an external  environment. Thus researchers are dedicating themselves to find efficient ways avoiding decoherence during quantum information process. Currently, the feasible methods include an active error-correction approach based on
the assumption that the most probable errors occur independently to a few qubits, which can be corrected via subsequent quantum operation \cite{bennett,chi,kosut,moussa,reed}, and alternative passive error-prevention scheme, where
the logical qubits are encoded into subspaces which do not
decohere because of symmetry \cite{lidar,beige,kempe,chen,pushin}. Recently, the function of dissipation is reexamined in Ref.~\cite{diehl,verstraete,vacanti,kastoryano,busch,shen,torre,rao,lin},
where the environment along can be used as a
resource to preparing entanglement and implementing universal quantum computing. In particular,
Kastoryano {\it et al.} consider a dissipative scheme for preparing
a maximally entangled state of two $\Lambda$-atoms in a high finesse
optical cavity \cite{kastoryano}, in which a pure steady singlet state is achieved with no need of state initialization.
\begin{figure}
\centering\scalebox{0.16}{\includegraphics{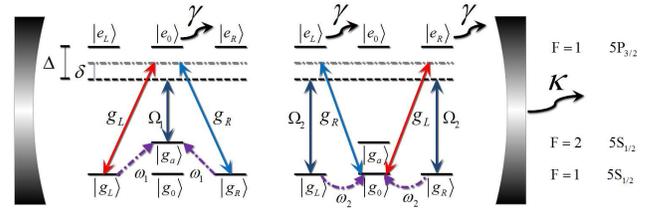} }
\caption{\label{p0}(Color online) Schematic view of the
configuration of atoms. The system consists of two $^{87}Rb$ atoms
simultaneously driven by optical lasers and microwave fields, coupled to a bi-mode cavity. The excited state of the first atom $|e_0\rangle$ can spontaneously decay into ground states $|g_L\rangle$, $|g_a\rangle$ and $|g_R\rangle$ with branching rate $\sqrt{\gamma_1/3}$, while the upper levels $|e_{L(R)}\rangle$ for the second atom is translated into $|g_{L(R)}\rangle$ and $|g_{0}\rangle$ with rate $\sqrt{\gamma_2/2}$, and we assume $\gamma_1=\gamma_2=\gamma$ throughout this manuscript. Also, the decay rates for the cavity modes are set to be the same $\kappa$.}
\end{figure}

Compared with other kinds of entanglement, high-dimensional entangled states have attracted
much interest, since it can enhance the violations
of local realism and the security of quantum cryptography. In the fields of linear optics,
two experiments utilize the spatial modes of the electromagnetic
field carrying orbital angular momentum to create high-dimensional
entanglement. In the context of
cavity quantum electrodynamics (QED), three-dimensional entanglement has also been realized in the unitary evolutionary dynamics based on resonant, and dispersive atom-cavity interactions \cite{mair,vaziri,zou,ye,lu,li}. In this paper, we put forward a dissipative method for preparing
a stationary three-dimensional entangled state. The motivation of our proposal is mainly based on the following truth: The typical decoherence factors in cavity QED system consist of atomic spontaneous emission and cavity decay, which have detrimental effects on schemes  based on unitary dynamics. However, the loss of cavity can used to stabilize a pure maximally entangled state when a suitable feedback control is applied \cite{1,2,3,shao}. Thus the spontaneous emission of atom becomes the only one detrimental factor. The result of our work shows that atomic spontaneous emission is able to
be a useful resource in respect of entanglement preparation, especially the fidelity of target state can even be better than the unitary evolution based schemes.

The structure of the manuscript is as follows. We derive the Lindblad master equation for preparation of three-dimensional entangled state with
 effective operator method in section II. We then
generalize the scheme to realization of an $N$-dimensional entangled state via introducing multi-level atoms and multi-mode cavity and discuss the effect of cavity decay on the fidelity in section III. This paper ends up with a conclusion in section IV.

\section{Effective master equation for open quantum systems}
We take into account a system composed by two $^{87}Rb$ atoms trapped in a bi-mode optical field, as shown in Fig.~\ref{p0}. The quantum states $|g_L\rangle$, $|g_0\rangle$, $|g_R\rangle$, and $|g_a\rangle$
correspond to atomic levels $|F=1,m_f=-1\rangle,$ $|F=1,m_f=0\rangle,$ $|F=1,m_f=1\rangle,$ and $|F=2,m_f=0\rangle$ of $5S_{1/2}$, and $|e_L\rangle$, $|e_0\rangle$, $|e_R\rangle$ correspond to $|F=1,m_f=-1\rangle,$ $|F=1,m_f=0\rangle,$ and $|F=2,m_f=1\rangle$ of $5P_{3/2}$. Without loss of generality, we apply two off-resonance $\pi$-polarized optical lasers, with Rabi frequencies $\Omega_{1(2)}$, detuning $\Delta$ to drive the transitions $|e_0\rangle\leftrightarrow|g_a\rangle$ for the first atom and
$|e_L\rangle\leftrightarrow|g_L\rangle$ and $|e_R\rangle\leftrightarrow|g_R\rangle$ for the second atom. The transition $|e_{0(R)}\rangle\leftrightarrow|g_{L(0)}\rangle$ and
$|e_{0(L)}\rangle\leftrightarrow|g_{R(0)}\rangle$ are coupled to the cavity modes $a_L$ and $a_R$ with coupling strength $g_L$ and $g_R$, detuning $\Delta-\delta$, respectively. In addition, two microwave fields with Rabi frequencies $\omega_1$ and $\omega_2$ are introduced to resonantly couple ground states as to acquire a steady-state entanglement during the dissipative process.

In a rotating frame, the master equation describing the interaction between quantum systems and external environment is in the Lindblad form
\begin{equation}\label{H1}
\dot{\hat{\rho}}=i[\hat{\rho},\hat{H}]+\sum_j\hat{L}_j\hat{\rho} \hat{L}_j^{\dag}-\frac{1}{2}(\hat{L}_j^{\dag}\hat{L}_j\hat{\rho}+\hat{\rho} \hat{L}_j^{\dag}\hat{L}_j).
\end{equation}
In connection with cavity QED, the Lindblad operator $L_j$ is closly related to two typical decoherence factors, i.e. the spontaneous emission rate $\gamma$ from the excited state of $^{87}Rb$ atom and
the leaky rate $\kappa$ of photon from the optical cavity. Thus the nine dissipation channels are denoted by
$L^{\gamma_1,g_{L(a,R)}}=\sqrt{\gamma/3}|g_{L,(a,R)}\rangle\langle e_0|$, $L^{\gamma_2,g_{L(0)}}=\sqrt{\gamma/2}|g_{L,(0)}\rangle\langle e_L|$,
$L^{\gamma_2,g_{R(0)}}=\sqrt{\gamma/2}|g_{R,(0)}\rangle\langle e_R|$, $L^{a_L}=\sqrt{\kappa}a_L$ and $L^{a_R}=\sqrt{\kappa}a_R$, respectively. For the sake of convenience, we have assumed a uniform dissipation rate for atoms and cavity modes. The Hamiltonian of the total system reads
\begin{eqnarray}\label{111}
\hat{H}&=&\hat{H}_0+\hat{H}_g+\hat{V}_++\hat{V}_-,\\
\hat{H}_0&=&\delta {\hat{a}}_L^{\dag}{\hat{a}}_L+\big[g_L(|g_L\rangle_{11}\langle e_0|+|g_0\rangle_{22}\langle e_R|){\hat{a}}_L^{\dag}+{\rm H.c.}\big]\nonumber\\
&&+\delta {\hat{a}}_R^{\dag}{\hat{a}}_R+\big[g_R(|g_R\rangle_{11}\langle e_0|+|g_0\rangle_{22}\langle e_L|){\hat{a}}_R^{\dag}+{\rm H.c.}\big]\nonumber\\
&&+\Delta(|e_0\rangle_{11}\langle e_0|+|e_L\rangle_{22}\langle e_L|+|e_R\rangle_{22}\langle e_R|),\\
\hat{H}_g&=&\omega_1(|g_L\rangle_{11}\langle g_a|+|g_R\rangle_{11}\langle g_a|)\nonumber\\&&+\omega_2(|g_L\rangle_{22}\langle g_0|+|g_R\rangle_{22}\langle g_0|)+{\rm H.c.},\\
\hat{V}_+&=&\hat{V}_-^{\dag}=\Omega_1(|e_0\rangle_{11}\langle g_a|)\nonumber\\&&+\Omega_2(|e_L\rangle_{22}\langle g_L|+|e_R\rangle_{22}\langle g_R|),
\end{eqnarray}
where $\hat{H}_0$ characterizes the strong interaction between atoms and quantized cavity fields, and $\hat{H}_g$ and
$\hat{V}_{\pm}$ correspond to the weakly driven fields of microwave and optical lasers, respectively. For simplicity, we set $g_L=g_R=g$, $\Omega_1=\Omega_2=\Omega$, and $\omega_1=-\omega_2=\omega$ in the following.
To gain a better insight into the effect of spontaneous emission  on the preparation of entanglement state, we first consider a perfect cavity without decay. According to the effective operator method \cite{reiter}, the excited states of the atoms and the cavity field modes can be adiabatically eliminated, provided that the Rabi frequency $\Omega$ of the optical pumping laser is sufficiently weak enough compared with $g$, $\delta$ and $\Delta$,  and the excited states
are not initially populated. Then we obtain the effective master equation as
\begin{eqnarray}\label{HHH}
\dot{\hat{\rho}}&=&i[\hat{\rho},\hat{H}_{\rm eff}]+\sum_j\hat{L}_{{\rm eff},j}\hat{\rho} \hat{L}_{{\rm eff},j}^{\dag}-\frac{1}{2}(\hat{L}_{{\rm eff},j}^{\dag}\hat{L}_{{\rm eff},j}\hat{\rho}\nonumber\\&&+\hat{\rho} \hat{L}_{{\rm eff},j}^{\dag}\hat{L}_{{\rm eff},j}),
\end{eqnarray}
where
\begin{eqnarray}\label{H7}
&&\hat{H}_{\rm eff}=-\frac{1}{2}[\hat{V}_-\hat{H}_{NH}^{-1}\hat{V}_++\hat{V}_-(\hat{H}_{NH}^{-1})^{\dag}\hat{V}_+]+\hat{H}_g,\nonumber\\
&&\hat{L}_{{\rm eff},j}=\hat{L}_j\hat{H}_{NH}^{-1}\hat{V}_+.
\end{eqnarray}
In the above expression, $\hat{H}_{NH}=\hat{H}_0-\frac{i}{2}\sum_j\hat{L}_j^{\dag}\hat{L}_j$ is a non-Hermitian Hamiltonian, and its inverted matrix can be written as $\hat{H}_{ NH}^{-1}=\hat{H}_{ NH_1}^{-1}+\hat{H}_{ NH_2}^{-1}+\hat{H}_{ NH_3}^{-1}$, explicitly
\begin{eqnarray}\label{H8}
\hat{H}_{ NH_1}^{-1}&=&\frac{g^2-3\delta\Delta^{'}}{9g^2\Delta^{'}-3\delta\Delta^{'2}}|X_1\rangle\langle X_1|
+\frac{1}{9}\bigg[\frac{8}{\Delta^{'}}-\frac{\delta}{3g^2-\delta\Delta^{'}}\bigg]\nonumber\\
&&|X_2\rangle\langle X_2|-\frac{1}{g^2-\delta\Delta^{'}}(\delta|X_3\rangle\langle X_3|+\Delta^{'}|-\rangle\langle -|)\nonumber\\
&&-\frac{\Delta^{'}}{3g^2-\delta\Delta^{'}}|+\rangle\langle+|+\bigg[\frac{2\sqrt{2}g^2}{9g^2\Delta^{'}-3\delta\Delta^{'2}}|X_2\rangle\langle X_1|\nonumber\\
&&+\frac{2\sqrt{2}g}{\sqrt{3}(3g^2-\delta\Delta^{'})}|+\rangle\langle X_1|-\frac{g}{\sqrt{3}(3g^2-\delta\Delta^{'})}\nonumber\\
&&|+\rangle\langle X_2|+\frac{g}{g^2-\delta\Delta^{'}}|-\rangle\langle X_3|+{\rm H.c.}\bigg],
\end{eqnarray}
\begin{eqnarray}\label{HHH}
\hat{H}_{ NH_2}^{-1}&=&-\frac{\delta}{2g^2-\delta\Delta^{'}}(|e_0g_L\rangle\langle e_0g_L|+|e_0g_R\rangle\langle e_0g_R|)\nonumber\\&&+\bigg[\frac{1}{\delta}-\frac{g^2}{2g^2\delta-\delta^{2}\Delta^{'}}\bigg]
(|g_Lg_L\rangle|1_L\rangle\langle g_Lg_L|\langle 1_L|\nonumber\\&&
+|g_Rg_L\rangle|1_R\rangle\langle g_Rg_L|\langle 1_R|+|g_Lg_R\rangle|1_L\rangle\langle g_Lg_R|\nonumber\\
&&\langle 1_L|+|g_Rg_R\rangle|1_R\rangle\langle g_Rg_R|\langle 1_R|)\nonumber\\&&
+\bigg\{\frac{g}{2g^2-\delta\Delta^{'}}[(|g_Lg_L\rangle|1_L\rangle+|g_Rg_L\rangle|1_R\rangle)\nonumber\\
&&\langle e_0g_L|+(|g_Lg_R\rangle|1_L\rangle+|g_Rg_R\rangle|1_R\rangle)\langle e_0g_R|]\nonumber\\
&&-\frac{g^2}{2g^2\delta-\delta^{2}\Delta^{'}}(|g_Rg_L\rangle|1_R\rangle\langle g_Lg_L|\langle 1_L|\nonumber\\
&&+|g_Rg_R\rangle|1_R\rangle\langle g_Lg_R|\langle 1_L|)+{\rm H.c.}\bigg\},
\end{eqnarray}
\begin{eqnarray}\label{HHH}
\hat{H}_{ NH_3}^{-1}&=&-\frac{\delta}{g^2-\delta\Delta^{'}}(|g_ae_L\rangle\langle g_ae_L|+
|g_ae_R\rangle\langle g_ae_R|\nonumber\\
&&+|g_Le_L\rangle\langle g_Le_L|+|g_Re_R\rangle\langle g_Re_R|)
\nonumber\\&&-\frac{\Delta^{'}}{g^2-\delta\Delta^{'}}(|g_ag_0\rangle|1_R\rangle\langle g_ag_0|\langle 1_R|\nonumber\\
&&+|g_ag_0\rangle|1_L\rangle\langle g_ag_0|\langle 1_L|+|g_Lg_0\rangle|1_R\rangle\langle g_Lg_0|\langle 1_R|\nonumber\\
&&+|g_Rg_0\rangle|1_L\rangle\langle g_Rg_0|\langle 1_L|)\nonumber\\
&&+\frac{g}{g^2-\delta\Delta^{'}}(
|g_ag_0\rangle|1_R\rangle\langle g_ae_L|\nonumber\\
&&+|g_ag_0\rangle|1_L\rangle\langle g_ae_R|+|g_Lg_0\rangle|1_R\rangle\langle g_Le_L|\nonumber\\&&+
|g_Rg_0\rangle|1_L\rangle\langle g_Re_R|+{\rm H.c.}),
\end{eqnarray}
where $\Delta^{'}=\Delta-\frac{i\gamma}{2}$  and the vacuum states of cavity modes are discarded and we have adopted the notation
$|X_1\rangle=\frac{1}{\sqrt{3}}(|g_Le_R\rangle+|g_Re_L\rangle+|e_0 g_a\rangle)$,
$|X_2\rangle=\frac{1}{\sqrt{6}}(|g_Le_R\rangle+|g_Re_L\rangle-2|e_0 g_a\rangle)$,
$|X_3\rangle=\frac{1}{\sqrt{2}}(|g_Le_R\rangle-|g_Re_L\rangle)$, and
$|+\rangle=\frac{1}{\sqrt{2}}(|g_Lg_0\rangle|1_L\rangle+|g_Rg_0\rangle|1_R\rangle)$,
$|-\rangle=\frac{1}{\sqrt{2}}(|g_Lg_0\rangle|1_L\rangle-|g_Rg_0\rangle|1_R\rangle)$.
On the basis of Eq.~(\ref{H7}), we have the effective Hamiltonian as
\begin{eqnarray}\label{HHH}
\hat{H}_{\rm eff}&=&\Omega^2 {\rm Re}\bigg[\frac{\delta}{g^2-\delta^{'}\Delta^{'}}+\frac{\delta}{2g^2-\delta^{'}\Delta^{'}}\bigg](|g_ag_L\rangle\langle g_ag_L|\nonumber\\
&&+|g_ag_R\rangle\langle g_ag_R|)+\Omega^2 {\rm Re}\bigg[\frac{\delta}{g^2-\delta\Delta^{'}}\bigg](|g_Lg_L\rangle\langle g_Lg_L|\nonumber\\&&+|g_Rg_R\rangle\langle g_Rg_R|+|T_3\rangle\langle T_3|)\nonumber\\&&-\Omega^2 {\rm Re}\bigg[\frac{g^2-3\delta\Delta^{'}}{9g^2\Delta^{'}-3\delta\Delta^{'2}}\bigg]|T_1\rangle\langle T_1|\nonumber\\&&-\Omega^2 {\rm Re}\bigg[\frac{2\sqrt{2}g^2}{9g^2\Delta^{'}-3\delta\Delta^{'2}}\bigg](|T_1\rangle\langle T_2|+{\rm H.c.})\nonumber\\
&&-\Omega^2 {\rm Re}\bigg\{\frac{1}{9}\bigg[\frac{8}{\Delta^{'}}-\frac{\delta}{3g^2-\delta\Delta^{'}}\bigg]\bigg\}|T_2\rangle\langle T_2|+\hat{H}_g,
\end{eqnarray}
where
$
|T_1\rangle=\frac{1}{\sqrt{3}}(|g_Lg_R\rangle+|g_Rg_L\rangle+|g_ag_0\rangle)
$
is the desired three-dimensional entangled state and
$|T_2\rangle=\frac{1}{\sqrt{6}}(|g_Lg_R\rangle+|g_Rg_L\rangle-2|g_ag_0\rangle)$,
$|T_3\rangle=\frac{1}{\sqrt{2}}(|g_Lg_R\rangle-|g_Rg_L\rangle)$. The effective Lindblad
operators induced by the spontaneous emission take the form of
\begin{eqnarray}\label{HHH}
\hat{L}_{\rm eff}^{\gamma_{1,g_{L(a,R)}}}&=&\frac{\Omega\sqrt{\gamma}}{\sqrt{3}}\bigg\{\frac{1}{\sqrt{3}}|g_{L(a,R)}
g_0\rangle\bigg[\bigg(\frac{g^2-3\delta\Delta^{'}}{9g^2\Delta^{'}-3\delta\Delta^{'2}}
\nonumber\\
&&-\frac{4g^2}{9g^2\Delta^{'}-3\delta\Delta^{'2}}\bigg)\langle T_1|+\frac{\sqrt{2}}{\sqrt{3}}\bigg(\frac{2g^2}{9g^2\Delta^{'}-3\delta\Delta^{'2}}
\nonumber\\
&&-\frac{8}{9\Delta^{'}}+\frac{\delta}{3g^2-\delta\Delta^{'}}\bigg)\langle T_2|\bigg]-\frac{\delta}{2g^2-\delta\Delta^{'}}\nonumber\\
&&(|g_{L(a,R)}g_L\rangle\langle g_ag_L|+|g_{L(a,R)}g_R\rangle\langle g_ag_R|)\bigg\},
\end{eqnarray}
\begin{eqnarray}\label{HHH}
\hat{L}_{\rm eff}^{\gamma_{2,g_{L(R)}}}&=&\frac{\Omega\sqrt{\gamma}}{\sqrt{2}}
\bigg\{\frac{1}{\sqrt{3}}|g_{R(L)}g_{L(R)}\rangle\bigg[\bigg(\frac{g^2-3\delta\Delta^{'}}{9g^2\Delta^{'}-3\delta\Delta^{'2}}
\nonumber\\
&&+\frac{2g^2}{9g^2\Delta^{'}-3\delta\Delta^{'2}}\bigg)\langle T_1|+\frac{\sqrt{2}}{\sqrt{3}}\bigg(\frac{2g^2}{9g^2\Delta^{'}-3\delta\Delta^{'2}}
\nonumber\\
&&+\frac{8}{9\Delta^{'}}-\frac{\delta}{3g^2-\delta\Delta^{'}}\bigg)\langle T_2|\bigg]-\frac{\delta}{g^2-\delta\Delta^{'}}\nonumber\\
&&(|g_ag_{L(R)}\rangle\langle g_ag_{L(R)}|+|g_{L(R)}g_{L(R)}\rangle\langle g_{L(R)}g_{L(R)}|\nonumber\\&&\mp\frac{1}{\sqrt{2}}|g_{R(L)}g_{L(R)}\rangle\langle T_3|)\bigg\},
\end{eqnarray}
\begin{eqnarray}\label{HHH}
\hat{L}_{\rm eff}^{\gamma_{2,g_{0}}}&=&\frac{\Omega\sqrt{\gamma}}{\sqrt{2}}\bigg\{\frac{1}{\sqrt{3}}(|g_Rg_0\rangle+|g_Lg_0\rangle)
\bigg[\bigg(\frac{g^2-3\delta\Delta^{'}}{9g^2\Delta^{'}-3\delta\Delta^{'2}}
\nonumber\\
&&+\frac{2g^2}{9g^2\Delta^{'}-3\delta\Delta^{'2}}\bigg)\langle T_1|+\frac{\sqrt{2}}{\sqrt{3}}\bigg(\frac{2g^2}{9g^2\Delta^{'}-3\delta\Delta^{'2}}
\nonumber\\
&&+\frac{8}{9\Delta^{'}}-\frac{\delta}{3g^2-\delta\Delta^{'}}\bigg)\langle T_2|\bigg]-\frac{\delta}{g^2-\delta\Delta^{'}}\big[|g_ag_0\rangle\nonumber\\
&&(\langle g_ag_L|+\langle g_ag_R|)+|g_Lg_0\rangle\langle g_Lg_L|\nonumber\\&&+|g_Rg_0\rangle\langle g_Rg_R|-\frac{1}{\sqrt{2}}(|g_Rg_0\rangle\nonumber\\
&&-|g_Lg_0\rangle)\langle T_3|\big]\bigg\}.
\end{eqnarray}
In order to have a compact form for the above expression, we have employed $|g_Lg_R\rangle$, $|g_Rg_L\rangle$, and $|g_ag_0\rangle$ to represent the {\it ket}s instead of $|T_{1(2,3)}\rangle$. It is worth noting that if we set the cavity detuning $\delta$ from two photon resonance meeting
the requirements $\delta=g^2/\Delta$, $\Delta\gg \gamma$, the other decay rates approximately equal to zero except the following dominant parts
\begin{figure}
\scalebox{0.26}{\includegraphics{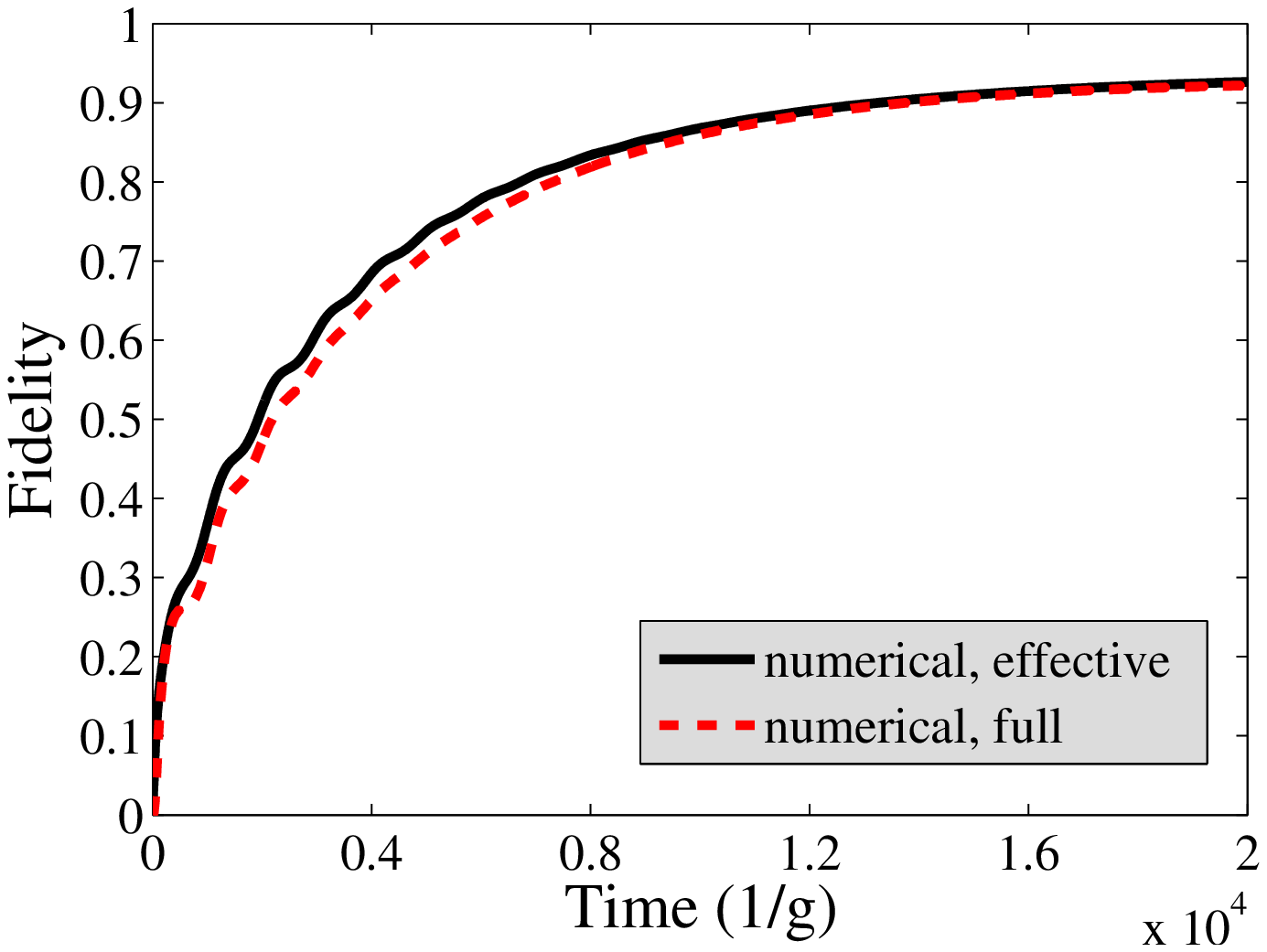} }
\scalebox{0.26}{\includegraphics{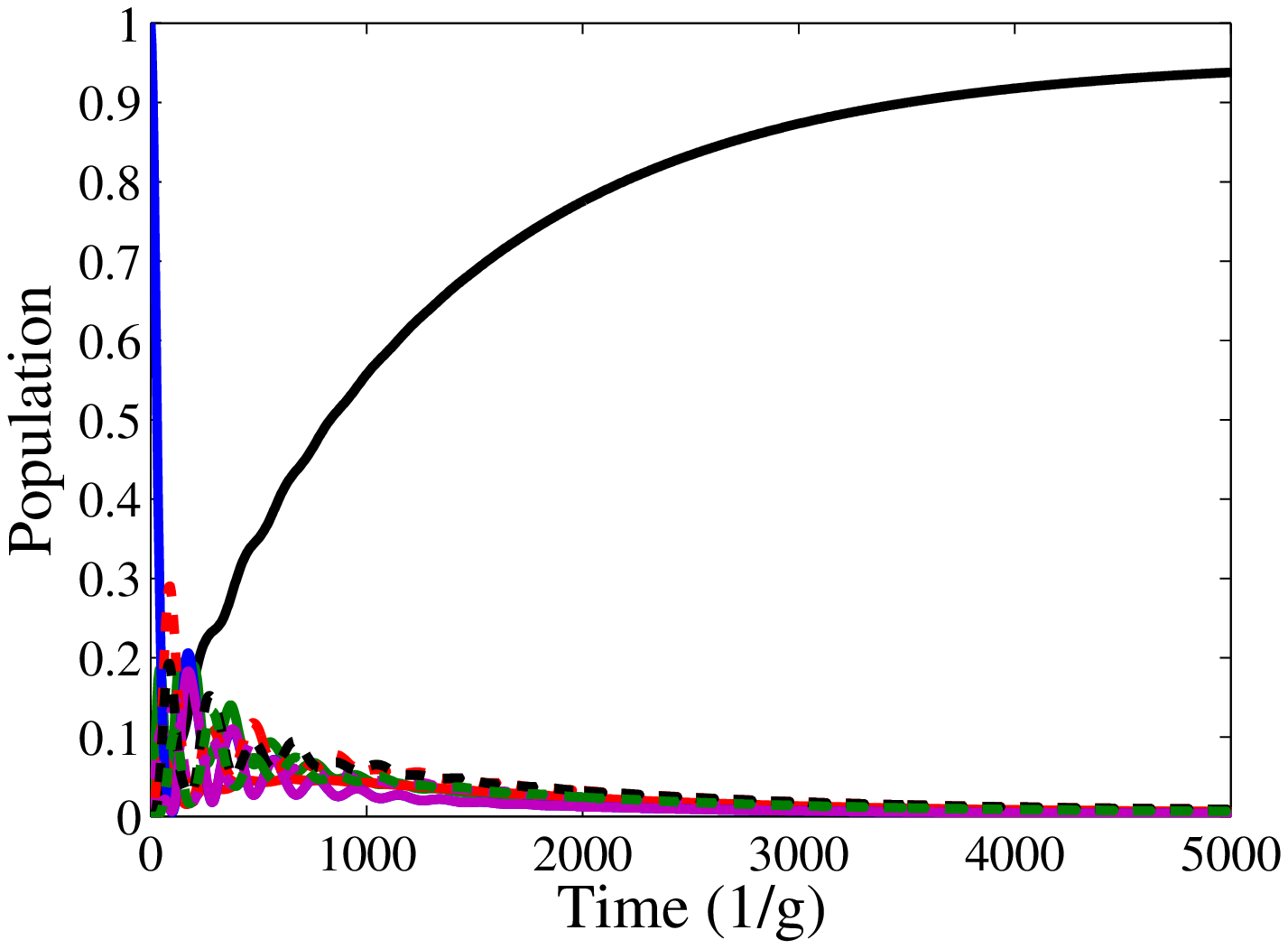} }
\caption{\label{p1}(Color online) Left panel: The comparison
of fidelities for preparation of the three-dimensional entangled
state $|T_1\rangle$ from an initial state $|g_ag_L\rangle$ with the full master equation (red dashed curve) and the effective one (black curve) under the given parameters $\Omega=0.02g$, $\omega=0.1\Omega$, $\Delta=g$, and $\kappa=0,\gamma=0.1g$.
Right panel: The populations of quantum states with optimized parameters
$\Omega=0.03g$, $\omega=0.4\Omega$, and $\Delta=g$ to achieve a stationary state within a short time corresponding to the same dissipation rate $\kappa=0,\gamma=0.1g$.}
\end{figure}
\begin{eqnarray}\label{H15}
\hat{L}_{\rm eff}^{\gamma_{2,g_{L(R)}}}&=&\frac{\sqrt{\gamma}}{\sqrt{2}}\frac{g_{\rm eff}}{\delta\gamma
/(2\Delta)}\bigg[\bigg(\frac{1}{2}|T_3\rangle\mp\frac{1}{\sqrt{6}}
|T_1\rangle\mp\frac{1}{2\sqrt{3}}
|T_2\rangle\bigg)\nonumber\\
&&\langle T_3|+|g_{L(R)}g_{L(R)}\rangle\langle g_{L(R)}g_{L(R)}|\nonumber\\&&+|g_ag_{L(R)}\rangle\langle g_ag_{L(R)}|\bigg],
\end{eqnarray}
\begin{eqnarray}\label{H16}
\hat{L}_{\rm eff}^{\gamma_{2,g_{0}}}&=&\frac{\sqrt{\gamma}}{\sqrt{2}}\frac{g_{\rm eff}}{\delta\gamma
/(2\Delta)}\bigg[\bigg(\frac{1}{\sqrt{3}}
|T_1\rangle-\frac{2}{\sqrt{6}}|T_2\rangle\bigg)(\langle g_ag_L|\nonumber\\
&&+\langle g_ag_R|)+|g_Lg_0\rangle\langle g_Lg_L|+|g_Rg_0\rangle\langle g_Rg_R|\nonumber\\&&-\frac{1}{\sqrt{2}}(|g_Rg_0\rangle-|g_Lg_0\rangle)\langle T_3|\bigg].
\end{eqnarray}
where $g_{\rm eff}=g\Omega/\Delta$. The application of microwave fields is crucial to our scheme, because it guarantees $|T_1\rangle$ remains the dark state while
 other ground states are coupled to each other. Therefore, the three-dimensional entangled state $|T_1\rangle$ is able to be achieved from an arbitrary initial state via the effective dissipation induced by spontaneous emission. In the left panel of Fig.~\ref{p1}, we plot the fidelities $F(|T_1\rangle,\hat{\rho})=\langle T_1|\hat{\rho}|T_1\rangle$ for creation of $|T_1\rangle$ with the full  and the effective master equations, from which we see that under the given parameters the full and the effective
dynamics of the system are in excellent agreement. In the right panel, we further optimize the parameters to make the entangled state reach stable in a shorter time.

\section{Generalization to high-dimensional entangled state}

\begin{figure}
\centering\scalebox{0.21}{\includegraphics{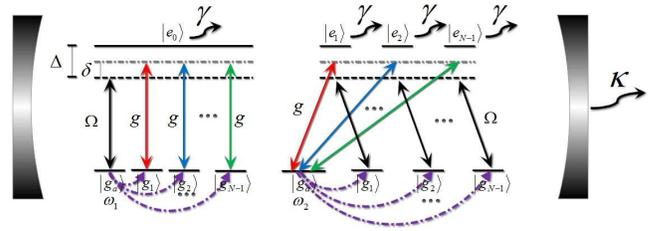} }
\caption{\label{px}(Color online) A potential atomic energy-level diagram to be used for generating an $N$-dimensional entangled state
$1/\sqrt{N}(|g_ag_a\rangle+|g_1g_1\rangle+|g_2g_2\rangle+\cdots+|g_{N-1}g_{N-1}\rangle)$.}
\end{figure}

\begin{figure}
\scalebox{0.25}{\includegraphics{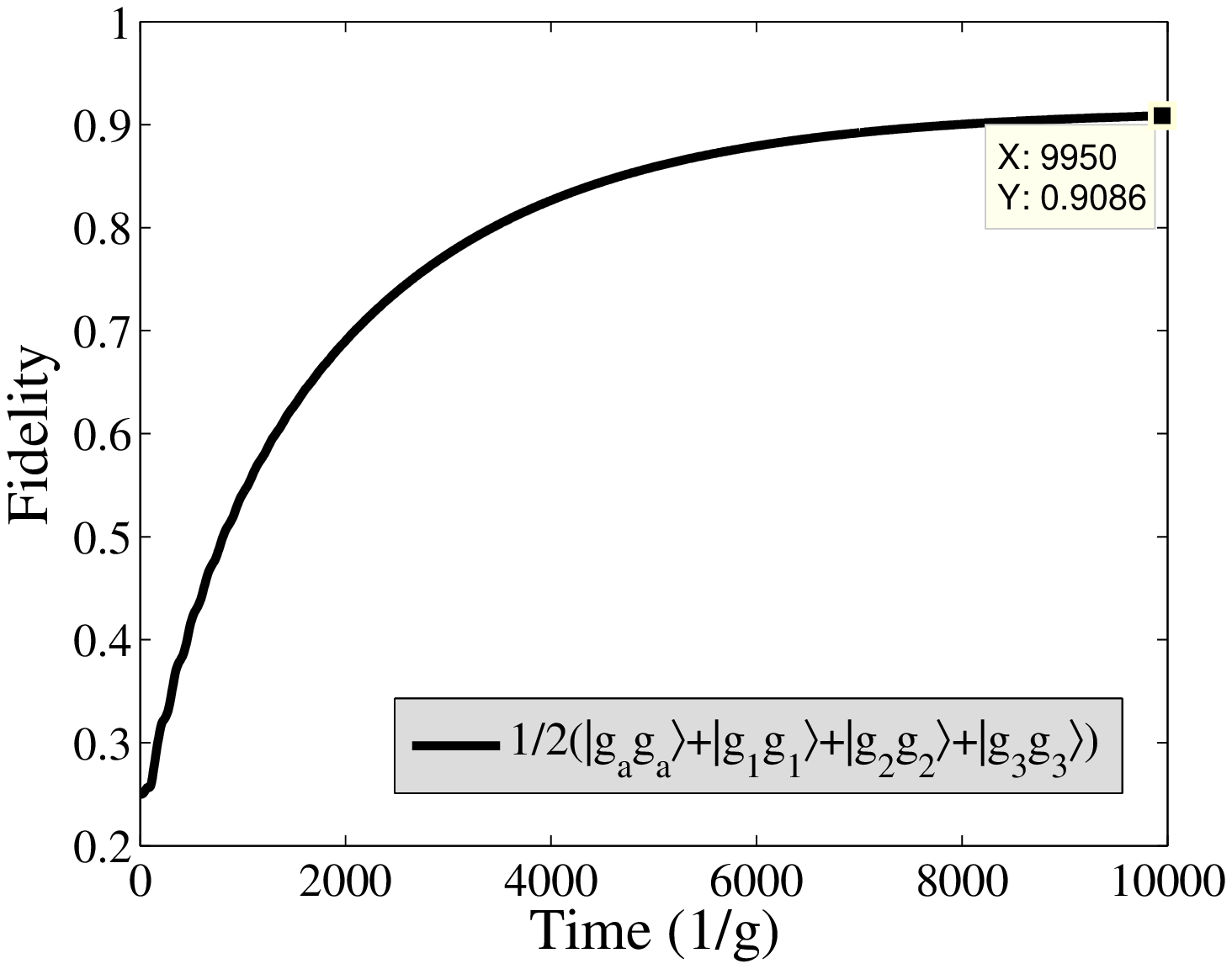} }
\scalebox{0.26}{\includegraphics{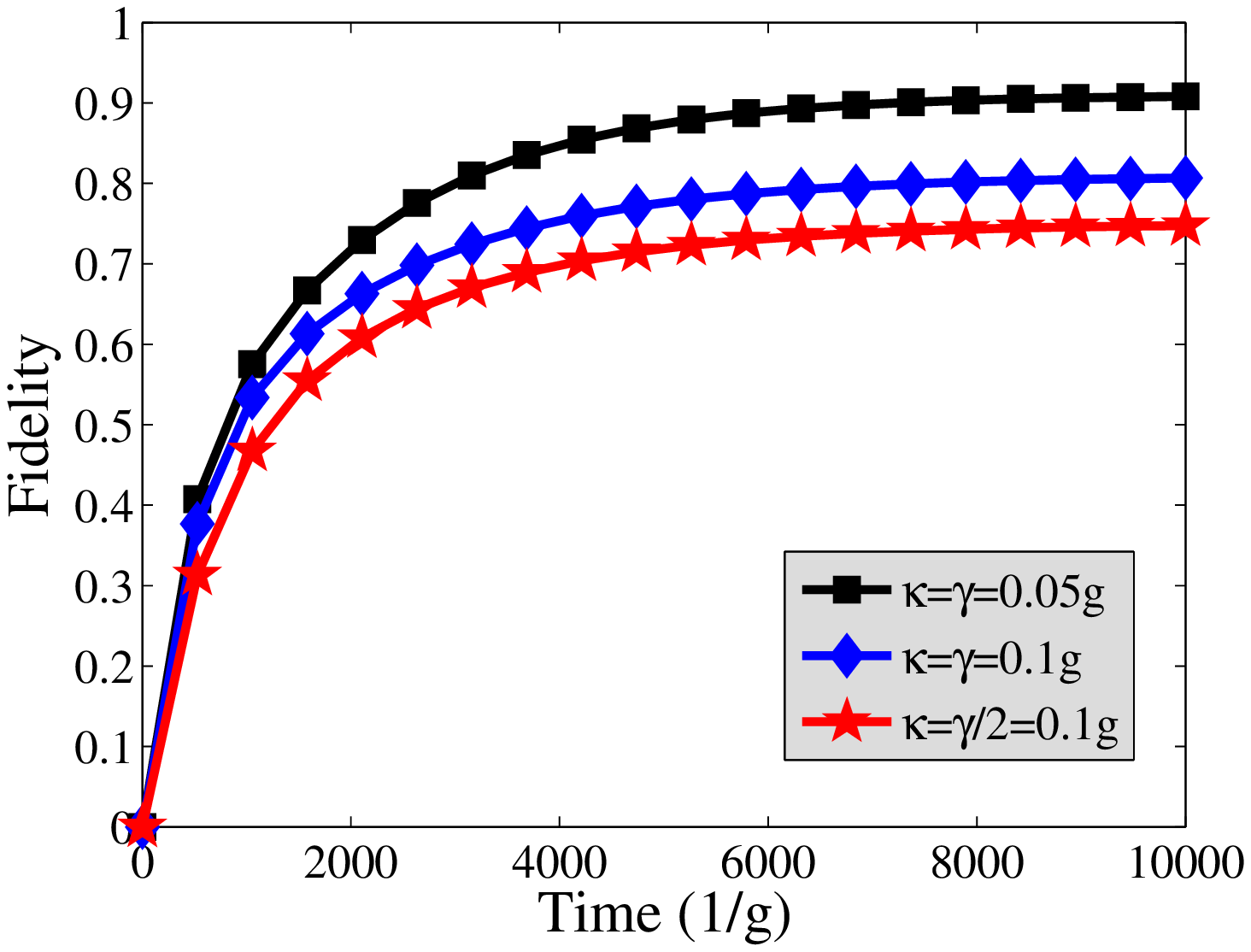} }
\caption{\label{py}(Color online) Left panel: The fidelity for preparation of four-dimensional entangled state with the same parameters shown in the right panel of Fig.~\ref{p1}, the initial state is randomly chosen as $|g_ag_a\rangle$. Right panel: The effect of cavity loss on the preparation of three-dimensional entangled state. }
\end{figure}

The successful use of dissipation to deterministic creation of three-dimensional entangled state mainly relies on the
effective level structure of atoms, i.e. we require transitions from a common excited (ground) state of first (second) atom to two ground (excited) states coupled by
two orthogonal cavity modes, while other transitions are driven by off-resonance optical lasers. Thus it is possible to generalize our model to
prepare high-dimensional entangled state if we design the atomic energy-level diagram following the similar rules. In Fig.~\ref{px}, we suppose
two potential multi-level atoms strongly interact with a multi-mode optical cavity, which is a direct extension of Fig.~\ref{p0}. By introducing microwave fields that drive the transitions  $|g_0\rangle\leftrightarrow|g_{i}\rangle$ where $i=1,\cdots,N-1$, an $N$-dimensional entangled state
$1/\sqrt{N}(|g_ag_a\rangle+|g_1g_1\rangle+|g_2g_2\rangle+\cdots+|g_{N-1}g_{N-1}\rangle)$ will be carried out via spontaneous emission. In confirmation of our assumption, we numerically simulation of the fidelity for generating the four-dimensional entangled state with the full master equation in the left panel of Fig.~\ref{py}. Compared with the case of three-dimensional entangled state, a longer time is needed to stabilize the target state above the fidelity $90\%$. Hence it is not difficult to conclude that the increase of dimension is at the cost of convergence time.

\begin{figure}
\scalebox{0.5}{\includegraphics{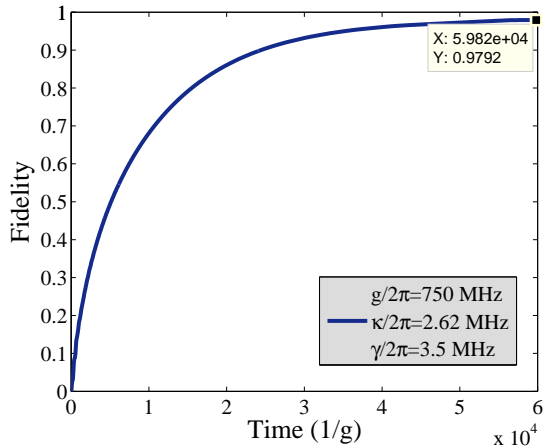} }
\caption{\label{pz}(Color online) Fidelity for generation of three-dimensional entangled state using an experimental cavity parameters. }
\end{figure}

Now we briefly discuss the effect of cavity decay on the performance for entanglement preparation.
In the right panel of Fig.~\ref{py}, we plot the fidelity by numerically solving the full master equation of Eq.~(\ref{H1}) incorporating  $\kappa$, three curves correspond to
different parameters of dissipation, i.e. $\kappa=\gamma=0.05g$, $\kappa=\gamma=0.1g$ and $\kappa=\gamma/2=0.1g$. The decrease of population for
$|T_1\rangle$ undoubtedly accompanied by a increase of population for other state. As the system approach to equilibrium, we will obtain a steady-mixed entanglement state. For certain cavity setup, the coupling strength between atom and cavity $g$, the cavity leakage rate $\kappa$, and the spontaneous emission rate $\gamma$ are fixed, thus we are allowed to modulate other parameters to achieve a three-dimensional entangled state with a relatively high fidelity. Fig~\ref{pz} illustrates the evolution of fidelity versus time with cavity parameters extracted from a recent experiment $(g, \kappa, \gamma)\sim 2\pi\times(750,2.62,3.5)$MHz \cite{spi}. A selection of $\Omega=0.02g,\omega=0.4\Omega,\Delta=g$ will lead to a fidelity about $98\%$, which overwhelms with  the value based on the unitary dynamics \cite{ye,lu,li}.

\section{Conclusion}

In conclusion, we have achieved a stationary three-dimensional entangled state via using the dissipation caused by spontaneous emission of atoms. The numerical simulation reveals the theory for effective operator agrees well with the full master equation under given parameters. This proposal is then extended to realize the $N$-dimensional entangled state in theory by considering two multi-level atoms interacting with a multi-mode cavity, which is confirmed by the simulation of implementing a four-dimensional entangled state.
The cavity decay plays a negative role on the state preparation, thus corresponding to different experimental situations, we need to regulate the Rabi frequencies of  both optical and microwave fields accurately so as to obtain a relatively high fidelity. We believe that
our work will be useful for the experimental realization of quantum
information in the near future.
\begin{center}{\bf{ACKNOWLEDGMENT}}
\end{center}

This work is supported by Fundamental Research Funds for the Central Universities under Grant No. 12SSXM001,  National
Natural Science Foundation of China under Grant Nos. 11204028 and 11175044, and National Research
Foundation and Ministry of Education, Singapore (Grant
No. WBS: R-710-000-008-271). X. Q. Shao was also supported in part by the Government of China through CSC.
\

\end{document}